# Simultaneous assessment of energy, charge state and angular distribution for medium energy ions interacting with ultra-thin self-supporting targets: a time-of-flight approach

R. Holeňák[1*], S. Lohmann[1], F. Sekula[1,2] and D. Primetzhofer[1]

[1]Department of Physics and Astronomy, Uppsala University, Box 516, S-751 20 Uppsala, Sweden

[2]Department of Physics and Nanotechnology, Brno University of Technology, Brno, Czech Republic

**Abstract**

We demonstrate simultaneous measurements of the charge state, energy and angular distribution of keV ions in transmission experiments through self-supporting foils. Using a time-of-flight approach we have introduced an electrostatic deflection apparatus as an extension to existing medium energy ion scattering (MEIS) instrumentation. Different positive, neutral and negative charge states have been discriminated and quantified for initially singly charged beams of He, N, O and Ne in the energy range from 25 to 250 keV. In parallel, the ion energy after interaction with the target has been assessed for all detected particles, while particles can be discriminated by deflection angle. Self-supporting thin carbon foils were used as samples to benchmark our experiments with literature data where available.

*Corresponding author: radek.holenak@physics.uu.se

The energy deposition by keV ions in materials is of ultimate importance for materials modification processes, ranging from space weathering over sputter deposition to ion implantation [1–3]. The observed energy deposition into the electronic structure of the target material is caused by a number of inelastic interactions: Apart from direct excitations in binary collisions, the charge state of the projectile can fluctuate as a consequence of electron loss and capture in close collisions with the target atoms. These ionization processes not only directly translate into an electronic energy loss but also alter the interaction probability of the ionic projectile with the electronic system. Typical time scales for these charge exchange processes are expected to be on the order of femtoseconds or below [4]. For keV ions, these fast processes are the basis for the analytical power of surface analytical tools like low-energy ion scattering [5]. In this energy range, charge exchange is expected to significantly contribute to the total energy deposition into the electronic system for ions other than protons [6,7]. Accurate knowledge of the charge state distribution of ions emerging from a target can thus shed light on the mechanisms underlying ion–solid interactions, helping to improve our understanding of ultrafast processes which trigger sputtering [8] as well as photon and electron emission [9–11]. In this context, highly charged ions with keV energies have been employed as a probe [12] showing e.g. the apparent similarity of neutralization in materials with such different properties as graphene and $MoS_2$ [13]. The combination of the time-of-flight approach and use of large detectors allows for the emission products to be studied in coincidence with the ions [14], and can thus provide a better understanding of the effects of electron dynamics observable in ion-solid interactions [15]. Also from an applied perspective, studying charge exchange processes in the medium energy regime, in particular for more conventional ions, such as He or Ne, has recently gained additional relevance: in the emerging field of ion microscopy the projectile charge affects the detected secondary electron yields and, simultaneously, trajectory dependent energy loss in crystals allows for contrast in imaging [16,17].

In the absence of a comprehensive theory for the equilibrium charge-state distribution, several empirical formulas have been proposed. A widely used formula by Nikolaev and Dmitriev [18], later improved by Schiwietz and Grande [19], is, however, applicable only above the Bohr velocity (2.18 x $10^6$ m/s). In the studied energy regime only a few monolayers of surface contamination consisting predominantly of carbon are enough to redefine the equilibrium charge state previously established in the sample bulk [20]. For that reason, the most extensive experimental work on charge fractions was performed using carbon foils. Almost identical compositions of surface and bulk ensures that the detected exit charge state distribution corresponds to the equilibrium charge state of carbon. For decades, self-supporting carbon foils found their main utilization in research applications as a stripping medium. A strong motivation for studying this specific system turned out to be the development of time-of-flight spectrometers and tandem accelerators [21–24]. An extensive contribution to the literature on charge fractions at low energies, around and way below Bohr velocity, was done by various space research groups [25–30], also developing new models applicable in this region [31]. There exists, however, a significant gap at intermediate ion energies (20-200 keV). This gap renders the transition regime from mainly neutral projectiles leaving the target materials to higher charge states being dominant largely unexplored.

In this work, we demonstrate a setup for charge state and simultaneous energy discrimination of ions transmitted through thin targets in a system capable to produce beams ranging from approximately 10 keV protons to multiply charged ions at several hundred keV with a wide choice of ion species. The introduced deflection apparatus serves as an extension to an existing medium energy ion scattering (MEIS) system increasing the information extractable in transmission experiments. The employed time-of-flight approach providing (sub-)nanosecond pulses allows for direct simultaneous measurement of all charge fractions as well as the energy loss of the projectiles transmitted through self-supporting foils. The performance of the equipment is benchmarked by existing literature data from carbon and new data are obtained in the medium energy range.

A deflection unit was installed into the existing TOF-MEIS setup at Uppsala University [32,33]. The experimental chamber features a 6-axis goniometer and a large, position-sensitive microchannel plate detector with diameter 120 mm from RoentDek [34]. The detector can be rotated freely around the centre of the scattering chamber and covers a solid angle of 0.13 sr. For the present experiments, the detector is positioned 290 mm behind the sample. The position of detected particles is determined with help of two perpendicular delay lines [35]. The incident single-charge ion beam of He, N, O or Ne passes through a set of apertures and is collimated to a beam spot size smaller than 1x1 mm$^2$ and beam angular divergence better than 0.056° when striking the carbon foil. Before reaching the experimental chamber, electrostatic chopping combined with a gating pulse shapes the incident beam into pulses with a duration of 1-2 ns while maintaining the steady-state current impinging on the foil at 2-3 fA. The operating pressure in the experimental chamber is 5 x 10$^{-8}$ mbar.

Behind the foil, transmitted particles are collimated by a narrow vertical slit that serves as an entrance aperture for an electrostatic deflection system. The electrostatic deflection apparatus consists of a parallel plate capacitor and entrance aperture and is attached to a L-shaped arm. A rotary feedthrough connection allows for freedom in positioning the apparatus in a continuous manner. Hence, the unit permits for fine inline adjustment when used for charge fraction measurement and can be rotated away without the need to break the vacuum, e.g. for alignment of single crystalline target materials or coincidence measurements [6,14]. A 3D model of the apparatus with schematic diagram of the experimental setup is shown in Fig. 1a.

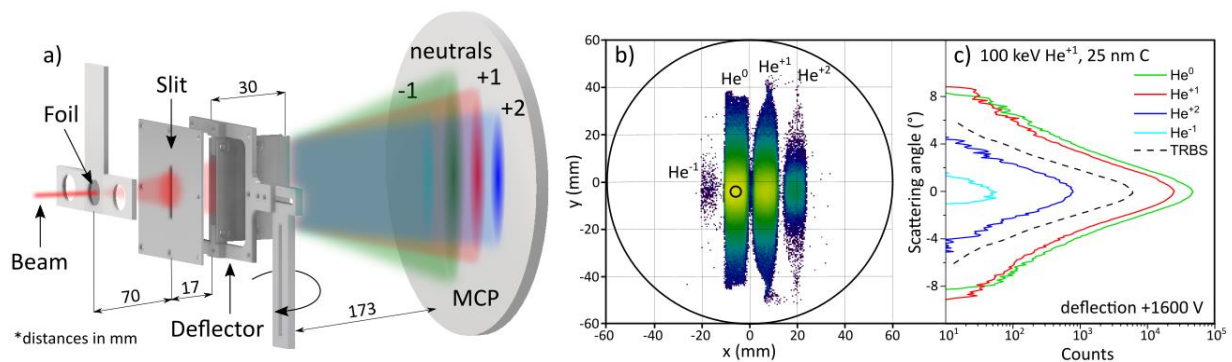

*Figure 1* a) Schematic view of the experimental setup. Particles transmitted through the target are collimated by a slit and discriminated according to their charge by capacitor plates. The impact position of each particle is recorded by a position-sensitive detector consisting of MCPs and delay lines. b) Detector image for an incident beam of He+ at 100 keV transmitted through a 25 nm thick carbon foil. Four vertical bands can be identified, corresponding to negative ions (leftmost), neutrals and, with

*increasing deflection in the x-direction, to singly and doubly charged ions. c) Angular distributions for each charge integrated over the respective band together with a profile simulated by TRBS.*

The geometry of the apparatus was designed to obtain good spatial separation of the different charge states for the whole range of projectiles and energies. A vertical slit (5mm x 20mm), slightly off-centre and 1 mm closer to the biased plate, serves as the aperture enhancing the spatial separation of neighbouring charges and simultaneously allowing for studying the angular distribution of the transmitted beam in y-direction covering an acceptance angle of +/-8°. On the otherwise grounded apparatus, one of the deflection plates is biased through the external voltage source maintaining a field up to 500 V/mm.

Carbon foils with nominal thicknesses of 25 nm and 50 nm were stretched over holes in a metal carrier by the floating method described elsewhere [36] and attached to the goniometer. We have not carried out an independent measurement of foil thickness since only the knowledge of the final ion energy is relevant. In our approach, the particle energy after transmission through the sample is measured via its flight time. Steady count rate on the MCP after being reduced by the entrance slit was kept around 5000/s ensuring good statistic for all charges observed. In the range of studied energies and charge states, the response of the MCP detector was found to be uniform and reproducible.

An example of an image accumulated on the detector is shown in Fig. 1b for an incident beam of 100 keV $He^+$ through a 25 nm carbon foil. One can clearly distinguish four vertical bands corresponding to charge states -1, 0, +1, +2. Figure 1c shows the spatial distributions of charged particles in y-direction integrated over the band widths along with the distribution obtained from TRBS simulation [37]. For the present amorphous target system, the angular distributions in Fig. 1c show no significant dependence on the detected charge state indicating, if any, only a minor distorting effect of the electric field in y-direction. This condition is necessary to enable studies of the angular dependence of charge fractions e.g. for targets with long-range order. To obtain the charge-state distribution, the counts in each band are separately integrated. Sufficient statistics allows for choosing to integrate only projectiles which travelled through the foil along straight trajectories for comparison with literature data. A small rectangular area with extension 2 mm in x and 4 mm in y centred in each band to accommodate the intensity maximum was used for integration. Derived charge fractions were normalized to the sum of the counts integrated.

The measured exit charge fractions for H, N, O and Ne are plotted as a function of exit velocity in Fig. 2. Statistical uncertainties are, commonly, smaller than the data points and in average better than 1.3%. We also observe a fraction of negative ions for both helium and oxygen.

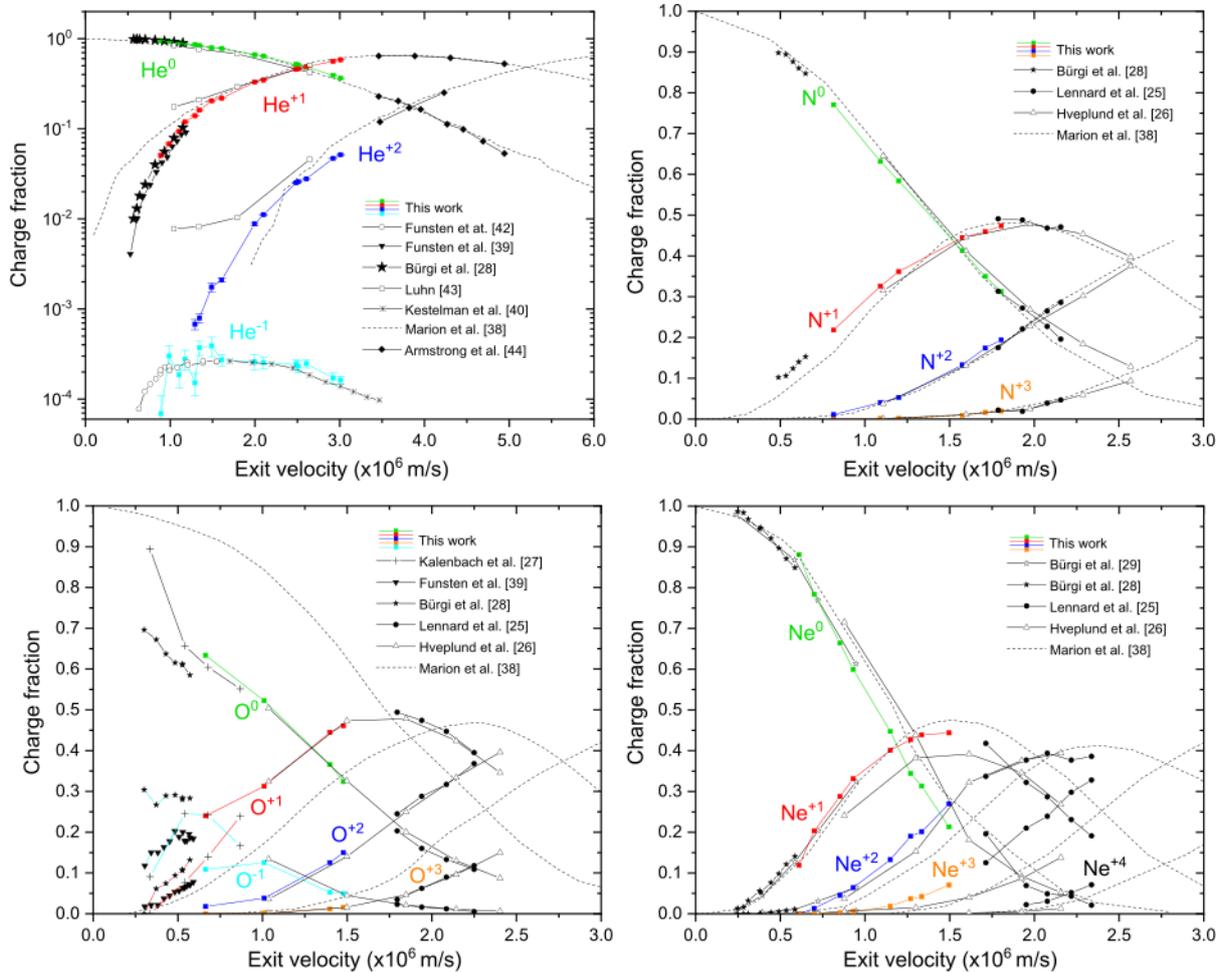

*Figure 2* Charge state fractions of helium, nitrogen, oxygen and neon after passing an amorphous carbon foil. The symbols indicate measurements from different authors. Our contribution is distinguished by colour. Dashed lines represent semiempirical calculations and extrapolations of existing data collected by Marion and Young [38].

The expected performance of the apparatus and potential aberrations induced by the electrical field were additionally examined by particle trajectory simulations using the SIMION software package [35]. While good homogeneity was achieved in between the parallel plates field lines outside the capacitor show an asymmetric behaviour. The potential effect of this inhomogeneity on particle trajectory and energy (due to a change in flight time) was simulated and considered in our evaluation. To avoid any possible effects of post-acceleration or deceleration induced by the field on the energy evaluation, only the detected neutrals were used for calculation of the velocities used in Fig. 2. From the band structure accumulated on the detector, the region around the centre of the neutrals fraction, marked with the black circle in Fig. 1b, was converted into energy spectra. The position of the energy peak directly provides the projectile's exit energy. The energy loss in the foil can then be easily calculated as the difference from the initial energy. The achieved energy resolution is generally better than 1 keV.

To address the effects of the field on charged particles emerging from the foil, we used the procedure described above to examine energy loss of the singly and doubly charged helium projectile with initial energy 100 keV transmitted through 25 nm thick carbon foil. The measured exit energy 92.2 keV of neutrals unaffected by any electric field was used in the simulation as initial energy for all charges before entering the apparatus. Fig. 3a shows the evolution of the particle energy as a function of the travelled distance from the foil to the detector. It can be seen that the inhomogeneity of the field leads to rapid deceleration of the projectiles in the space behind the grounded aperture towards the entrance of the deflector. When traversing in between the plates, the field separates the charged particles and consequently accelerates them on the way out. Upon arrival on the detector with a grounded grid in front, the projectiles regain their initial energy.

Even though the particle energy is finally conserved, the velocity variation along the trajectory leads to an increase of the absolute time-of-flight of the particle to the detector. Since the flight time is the experimentally accessible quantity in our approach, for given path length the time delay would indicate a difference in energy loss. The observed time delay is charge-state dependent and exceeds the time delay caused by the path prolongation due to the deflection of higher charges, when assuming constant velocity (see Fig. 3b). Experimentally observed time-of-flight values are reproduced by the SIMION simulation when considering the non-constant particle velocity. The error bars in the measured time-of-flight come from the uncertainty of the projectile's time peak position. In the simulation the error was estimated by moving the particle source by 1 mm to each side within the projected slit width.

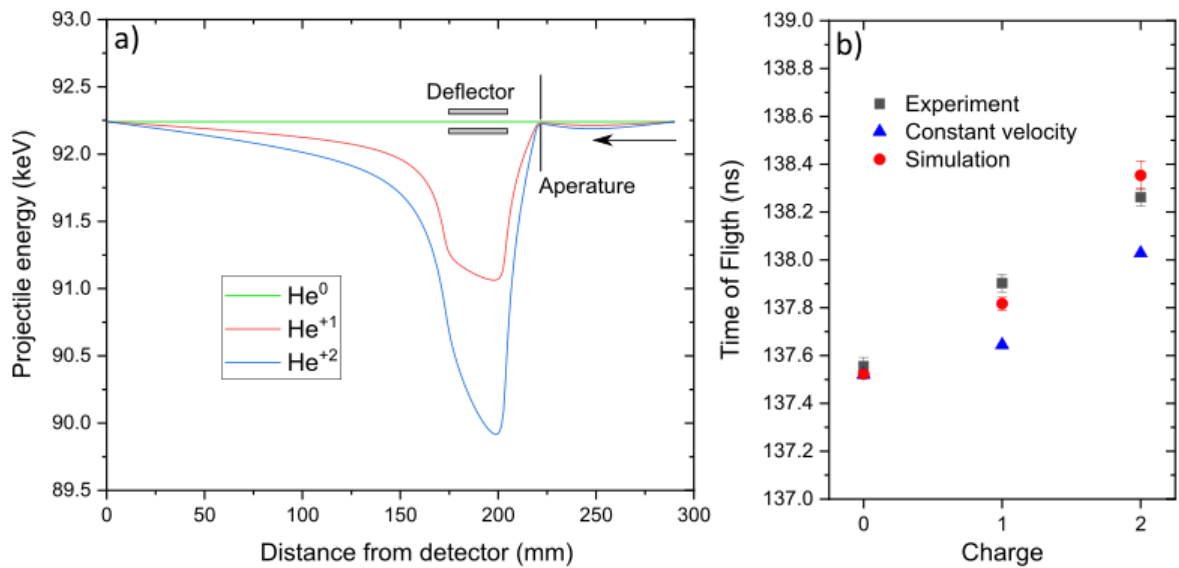

***Figure 3*** *a) Evolution of the kinetic energy of the projectiles along the flight path. Particles are travelling from right to left. b) Comparison of experimental values for the time-of-flight with the particle trajectory calculations.*

Along with increasing flight time of the projectiles, SIMION simulations predict a field-induced distortion of trajectories close to the maximum acceptance angle. This effect could be indeed observed in the experimental data when comparing band shapes as shown in Fig. 1b. For cases in which distributions were narrow enough to fit the acceptance angle defined by the aperture, the angular distributions could be compared with the profile expected from TRBS simulation. Within the acceptance angle of +/-8° the distributions for all charges were found identical, and in agreement with TRBS prediction. Our results are thus in accordance with previous studies where no correlation between exiting charge state and energy loss or deflection angle was observed [29,39].

Our data for all studied projectiles in the low energy region follow the data from Bürgi et al. [28,29] and the higher energies studied connect to measurements of Lenard et al. [25]. For nitrogen and oxygen, the data in the intermediate region, close to the maximum of charge +1, are overlapping with data from Hveplund et al. [26] which were one of the first experiments of this kind. However, in the case of neon, their data show an overall lower number of charged particles exiting the foil. The semi-empirical models plotted were, when proposed, lacking existing experimental data measured at low ion energies. This fact is especially apparent for oxygen, where a substantial fraction of projectiles is exiting the foil with a negative charge, different from predictions. The formation of negative ions has been studied earlier for different projectiles [40–42]. While our negative fractions for helium are in agreement with literature data, the results for oxygen scatter for all available data. In comparison to previous works, we report the lowest negative fraction.

The widest contribution of our experimental data to the charge fraction database in terms of energy range covered is for helium. In the intermediate region for projectile energies below 0.2 MeV, the only available data is of Luhn [43], who in his thesis also adopted the time-of-flight approach, but used a biased foil target. We reach agreement at higher energies interpolating data of Armstrong et al. [44]. Luhn's data are, however, at lower energies deviating from the trend given by data of Bürgi et al. where our data exhibits a smooth connection. Our dataset for charge +2, is found lower by one order of magnitude than the only available dataset of Luhn.

Despite the instrument's simplicity, its performance has been demonstrated by yielding trustworthy results in agreement with the available literature data. Moreover, for all investigated ion species, we could bridge the gap from existing low-energy data to higher energies by providing datasets covering all detectable charge states in a single study. Due to the wide acceptance angle in one direction, our system offers the possibility to study charge-exchange processes on more complex systems, including various 2D materials and single crystalline thin films. This capability is expected to yield new insights on the entanglement of charge exchange, energy deposition and elastic scattering

Accelerator operation is supported by the Swedish Research Council VR-RFI (Contracts No. 2017-00646_9) and the Swedish Foundation for Strategic Research (Contract No. RIF14-0053).